\documentclass[twocolumn,showpacs,amsmath,amssymb]{revtex4}

 \usepackage[colorlinks=true, linkcolor=blue, urlcolor=blue, citecolor=blue, anchorcolor=blue]{hyperref}
\usepackage{graphicx,amsmath,amssymb,bm,multirow}
\usepackage{amsthm}
\usepackage{amsfonts}
\usepackage{dcolumn}
\usepackage{bm}
\usepackage{mathrsfs}
\usepackage[usenames,dvipsnames]{color}
\newcommand{\bsub}{  \begin{subequations}}
\newcommand{\esub}{ \end{subequations}}

\newcommand{\jmy}[1]{\textcolor{black}{#1}}

\begin{document}
\title{Anatomy of molecular structures in $^{20}$Ne}
\author{E. F. Zhou$^{1}$, J. M. Yao$^{1,2,3}$\footnote{Corresponding author: jmyao@swu.edu.cn},
Z. P. Li$^{1}$, J. Meng$^{4,5,6}$, P. Ring$^{4,7}$}
\address{$^{1}$ School of Physical Science and Technology, Southwest University, Chongqing 400715, China}
\address{$^{2}$ Department of Physics, Tohoku University, Sendai 980-8578, Japan}
\address{$^{3}$ Department of Physics and Astronomy, University of North Carolina, Chapel Hill, North Carolina 27516-3255, USA}
\address{$^{4}$ State Key Laboratory of Nuclear Physics and Technology, School of Physics, Peking University, Beijing 100871, China}
\address{$^{5}$ School of Physics and Nuclear Energy Engineering, Beihang University, Beijing 100191, China}
\address{$^{6}$ Department of Physics, University of Stellenbosch, Stellenbosch 7602, South Africa}
\address{$^{7}$ Physik-Department der Technischen Universit\"at M\"unchen, D-85748 Garching, Germany}

\date{\today}

\begin{abstract}
We present a beyond mean-field study of clusters and molecular structures in low-spin states of $^{20}$Ne with a multireference relativistic energy density functional, where the dynamical correlation effects of symmetry restoration and quadrupole-octupole shapes fluctuation are taken into account with projections on parity, particle number and angular momentum in the framework of the generator coordinate method. Both the energy spectrum and the electric multipole transition strengths for low-lying parity-doublet bands are better reproduced after taking into account the dynamical octupole vibration effect. Consistent with the finding in previous studies, a rotation-induced dissolution of the $\alpha+^{16}$O molecular structure in $^{20}$Ne is predicted.
%\jmy{We conjecture that this peculiar phenomenon is partially attributed to the special deformation-dependent moment of inertia.}
%antisymmetrized molecular dynamics  and
\end{abstract}

\pacs{21.10.-k, 21.60.Jz, 21.10.Re}
% 21.60.Jz:   Nuclear Density Functional Theory and extensions
% 21.10.Re:  Collective levels
% 21.10.-k	Properties of nuclei; nuclear energy levels
\maketitle

The formation of clusters in nuclear many-body systems has continued being a very active field of research since it plays an important role in studies of nuclear structure, cluster decay, break-up reactions, and stellar nucleosynthesis~\cite{Ikeda80,Fynbo05,Oertzen06,Freer07,Kanada12}. For nuclear  matter at very low densities, light clusters can be formed and have a significant influence on the nuclear equation of state and therefore on the structure of neutron stars and for simulating supernova explosions~\cite{Typel10}. In finite nuclei, clusters are typically observed in excited states close to the corresponding decay threshold with a molecular structure, which in most cases is composed of one or several $\alpha$ particles with or without a closed-shell core surrounded  by several neutrons~\cite{Oertzen06,Freer07,Kanada12}. The molecular states in $\alpha$-conjugate stable nuclei with an equal, even number of protons and neutrons are well described by the so-called Ikeda diagram~\cite{Ikeda68,Ikeda80}. For neutron-rich nuclei, the excess neutrons form molecular orbits in the
low excitation energy region and tend to form atomic orbits around individual clusters in the region of higher excitation energy~\cite{Horiuchi12}.  In contrast to excited states,  the cluster structure in most of the ground states does not survive as well-separated $\alpha$-particles, but rather the cluster structure becomes more compact and the clusters overlap due to a stronger attractive inter-cluster interaction. As a result, their ground states have a ``duality nature"~\cite{Bayman58} or admixture of both mean-field and clustering ingredients, as indicated in the {\em ab initio} calculations of Refs.~\cite{Wiringa00,Chernykh07}. The transition of nuclear structure from mean-field to clustering structure under controlled parameters, such as nucleon number, excitation energy, temperature or angular momentum, has been of particular interest in nuclear physics research.

In the past decades, cluster structures in the deformed nucleus $^{20}$Ne have been extensively studied~\cite{Lee72,Matsuse75,Nemoto75,Marcos83,Kanada95,Taniguchi04,Kimura04,Ohta04} because this nucleus is one of a few nuclei with a strong admixture of cluster configurations in the ground state. In particular, it has a transitional character between deformed mean-field structures and cluster structures, as demonstrated in angular-momentum projected Hartree-Fock~\cite{Lee72}, resonating group method~\cite{Matsuse75}, $5\alpha$ generator coordinate method (GCM)~\cite{Nemoto75}, and anti-symmetrized molecular dynamics (AMD) calculations~\cite{Kanada95,Taniguchi04,Kimura04}, which predicted that the reflection-asymmetric $\alpha+^{16}$O molecular structure becomes weaker as the spin goes up in this way that the equilibrium distance between the two clusters is gradually decreasing. This phenomenon is in contradiction to the common sense that rotation elongates nuclear shapes due to the centrifugal force.
There should be a mechanism playing the role of anti-stretching and driving $^{20}$Ne to a state with an enhanced spin-orbit interaction dissolving the cluster structure~\cite{Kanada95,Kimura04}. However, to the best of our knowledge, the dynamical mechanism that drives the dissolution of  the $\alpha+^{16}$O molecular structure in the parity-doublet states of $^{20}$Ne is not clear and requires further investigations within a fully  microscopic approach capable of describing the coexistence of deformed mean-field and cluster structures.

In recent years, clustering phenomena in light $N=Z$ and neutron-rich nuclei were reexamined in the framework of energy density functional (EDF) methods~\cite{Ebran12}. It was claimed that the relativistic EDFs are characterized by deeper single-nucleon potentials and thus are prone to enhance the probability of forming cluster structures in both ground and excited states. A number of other EDF studies have shown that cluster structures appear in nuclear systems with high spin and large deformation~\cite{Arumugam05,Zhang10,Ichikawa11,Zhao15} or with low density~\cite{Girod13}. On one hand, all these studies demonstrate the feasibility of nuclear EDF theories  to explore the occurrence and evolution of $\alpha$-cluster structures. On the other hand, however, cluster structures built on a single-reference state could be unstable against shape fluctuations. A quantitative assessment of the effect of quantum fluctuations and a detailed description of the spectra require methods ``going beyond the mean field" by means of symmetry restoration and taking into account fluctuations in the nuclear shape. This framework is usually referred as generator coordinate method~\cite{Ring80}  or multireference EDF method~\cite{Duguet14}. Recently,  the stability of a linear four-$\alpha$ structure in high-lying excited states of $^{16}$O was examined with the multireference relativistic EDF method~\cite{Yao14-O}. In this paper, we extend this method by mixing quantum-number projected quadrupole-octupole shaped configurations in order to study molecular-like structures in the low-lying parity-doublet states of $^{20}$Ne.

The symmetry conserved wave function is constructed by superposing a set of quantum-number projected non-orthogonal mean-field reference states $\vert q\rangle$ around the equilibrium shape

\begin{equation}\label{gcmwf}
\vert J^\pi NZ; \alpha\rangle
=\sum_{\kappa\in\{q, K\}} f^{J\pi\alpha}_\kappa  \hat P^J_{MK} \hat P^N\hat P^Z  \hat P^\pi\vert q\rangle,
\end{equation}
where the generator coordinate $q$ stands for the discretized deformation parameters $\{\beta_2,  \beta_3\}$ of the reference states.  The $\hat P^G$s ($G\equiv J, \pi, N, Z )$ are projection operators onto angular momentum, parity, neutron and proton numbers, respectively~\cite{Ring80}. The weight functions $f^{J\pi\alpha}_\kappa$ are determined by the Hill-Wheeler-Griffin equation~\cite{Hill57},
\begin{equation}\label{HWG}
\sum_{\kappa_b} \left[  \mathscr{H}^{J\pi}_{\kappa_a, \kappa_b}-E^{J\pi}_{\alpha} \mathscr{N}^{J\pi}_{\kappa_a, \kappa_b}\right]f^{J\pi\alpha}_{\kappa_b}=0,
\end{equation}
where the Hamiltonian kernel $ \mathscr{H}^{J\pi}_{\kappa_a, \kappa_b}$ and the norm kernel
$\mathscr{N}^{J\pi}_{\kappa_a, \kappa_b}$ are given by,
\begin{equation}
\label{kernel}
 \mathscr{O}^{J\pi}_{\kappa_a, \kappa_b}
 =\langle q_a \vert  \hat O \hat P^J_{K_aK_b} \hat P^N\hat P^Z  \hat P^\pi\vert q_b\rangle
\end{equation}
with the operator $\hat O$ representing $\hat H$ and $1$, respectively and the index $\kappa$ for $\{q, K\}$.

\begin{figure}[t]
\includegraphics[width=10cm]{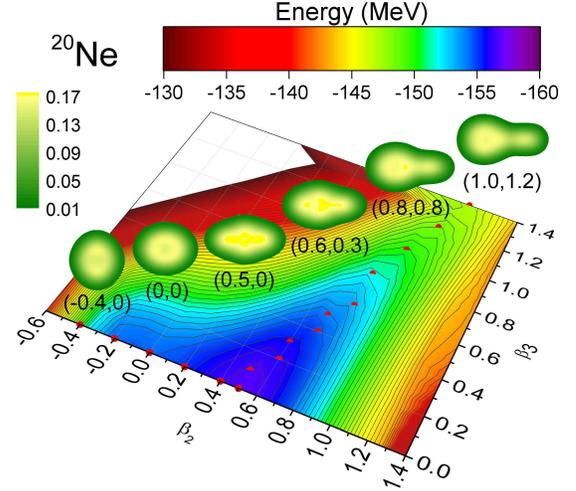}
\caption{(color online) The mean-field energy surface of $^{20}$Ne in two-dimensional $(\beta_2, \beta_3)$ deformation plane, where the optimal configurations (along the valley) are indicated with red dots. The density profiles of several selected configurations are plotted (in fm$^{-3}$).}
  \label{PES:MF}
\end{figure}

%%%%%%%%%%%%%%%%%%%%%%%%
The reference states $\vert q\rangle$ are generated from deformation constrained self-consistent relativistic mean-field calculations  based on the universal relativistic energy functional PC-PK1~\cite{Zhao10} through the variational principle
 \begin{equation}\label{Dirac}
 \delta \langle q\vert \hat H - \sum_{\tau=n, p} \lambda_\tau \hat N_\tau - \sum_{\lambda=1, 2,3} C_\lambda (\hat Q_{\lambda0} - q_\lambda)^2  \vert q\rangle=0
\end{equation}
 with the Lagrange multipliers $\lambda_\tau$ being determined by the constraints $\langle q \vert \hat N_\tau\vert q\rangle=N(Z)$.  The position of the center of mass coordinate is fixed at the origin to decouple the spurious states by the constraint $\langle q \vert \hat Q_{10} \vert q\rangle=0$. The $\hat N_\tau$ and $\hat Q_{\lambda 0}\equiv r^\lambda Y_{\lambda 0}$ are particle number and multipole moment operators. For the sake of simplicity, all the reference states are restricted to be axially deformed.  The deformation parameters $\beta_\lambda$ ($\lambda=2, 3$) are defined as
\begin{equation}\label{deformation}
  \beta_\lambda \equiv \dfrac{4\pi}{3A R^\lambda} q_\lambda, \quad R=1.2A^{1/3},
\end{equation}
with $A$ representing the mass number of the nucleus. The Dirac equation for  the nucleons is solved by expanding the Dirac spinors in a harmonic oscillator basis with ten shells~\cite{Gambhir90}.  Pairing correlations between nucleons are treated within the BCS approximation by using a density-independent $\delta$-force with a smooth cutoff~\cite{Krieger90}. The strength parameters of the pairing force are $V_n=-349.5$ and $V_p=-330.0$ MeV fm$^3$ for neutrons and protons, respectively. In the calculation of the projected kernels, the numbers of mesh points in the interval $[0,\pi]$ for the rotation angle $\beta$ and the gauge angle $\varphi$ are chosen as $N_\beta=14$ and $N_\varphi=7$, respectively. We adopt $N_\text{GCM}=54$ reference states covering the $(\beta_2,\beta_3\ge0)$ deformation plane in the GCM calculation. The convergence of the GCM calculation is checked by increasing or decreasing the number of configurations and by examining the behavior of the collective wave functions and energy dispersions~\cite{Bonche90}. Since the low-lying parity-doublet states of $^{20}$Ne are dominated by the deformed configurations with $(\beta_2>0,\beta_3)$, the oblate deformed configurations  ($\beta_2<0$) are excluded in our final GCM calculation to save computation time. Pfaffian techniques~\cite{Robledo09,Bertsch12} are implemented to avoid the sign problem for the norm overlaps. More details about the multireference relativistic energy density functional calculation have been discussed in Refs.~\cite{Yao10,Yao14,Yao15}.

The advantages of our method for studying the molecular structure in the low-lying parity doublets of $^{20}$Ne are as follows: 1) the strength of the spin-orbit interaction relevant for the formation of cluster structures is determined automatically by the derivative of the scalar and vector fields in all the reference states $\vert q\rangle$; 2) the fluctuations in quadrupole-octupole shapes relevant for the stability of cluster structures is taken into account; 3) the symmetry restorations relevant for spectroscopic studies of the evolution of cluster structures in excited states are implemented. This method provides a state-of-the-art calculation of nuclear low-lying parity-doublet states with possible quadrupole-octupole shapes.

\begin{figure}[tb]
\begin{center}
\includegraphics[width=8.5cm]{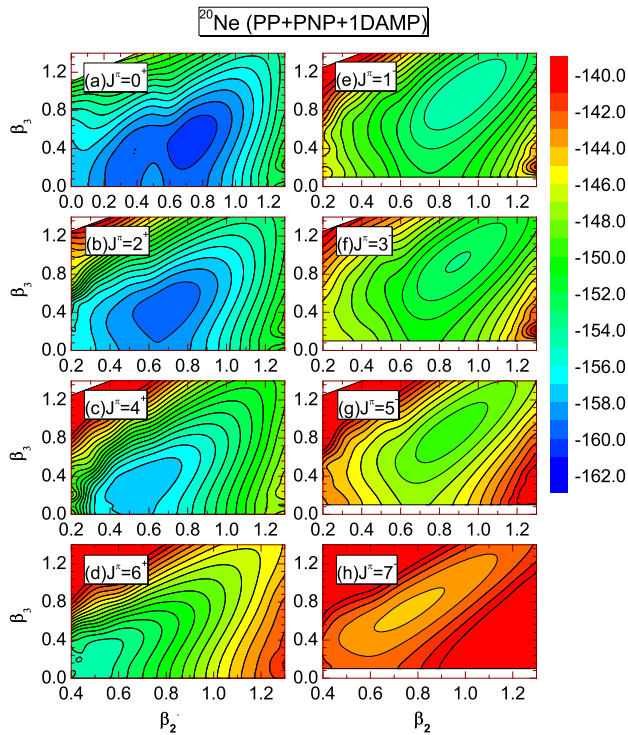}
\end{center}
\caption{(Color online) The energy surfaces of quantum-number projected states in $(\beta_2,\beta_3)$ deformation plane for $^{20}$Ne with projection onto spin-parity ($J^\pi$) and particle numbers ($N, Z$). Two neighboring contour lines are separated by 1.0 MeV.}
\label{PES:proj}
\end{figure}
\begin{figure}[tb]
\begin{center}
\includegraphics[width=7cm]{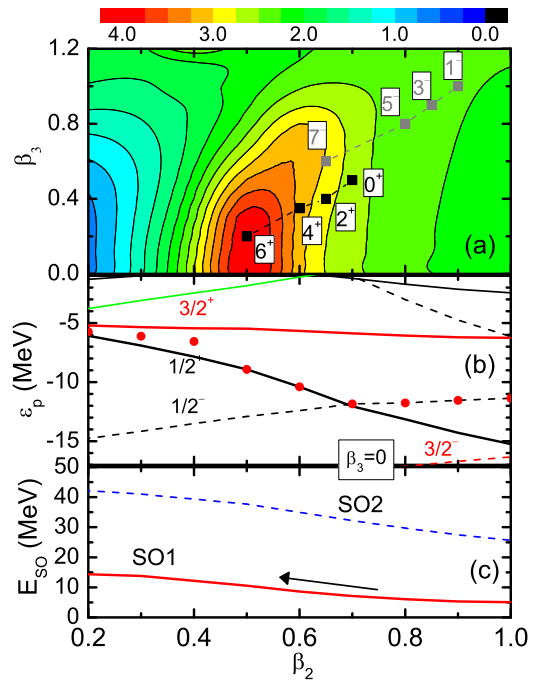} \vspace{-0.5cm}
\end{center}
\caption{(Color online) (a) The deformation-dependent moment of inertia ${\cal{J}}^{IB}_x$  ($\hbar^2$/MeV) along $x$-axis, cf.(\ref{MOI}), in $(\beta_2,\beta_3)$ plane, where the energy-minimum configuration on the projected energy surfaces with different spin-parity in Fig.\ref{PES:proj} are indicated. (b) Single-particle energy of protons in $^{20}$Ne as a function of deformation $\beta_2$($\beta_3=0$). Fermi energies are indicated with red dots. (c) Spin-orbit potential values ($SO1$ and $SO2$) of  interaction in the mean-field state as a function of $\beta_2$ ($\beta_3=0$).}
\label{fig:MoI}
\end{figure}

\begin{figure*}[]
\includegraphics[width=14cm]{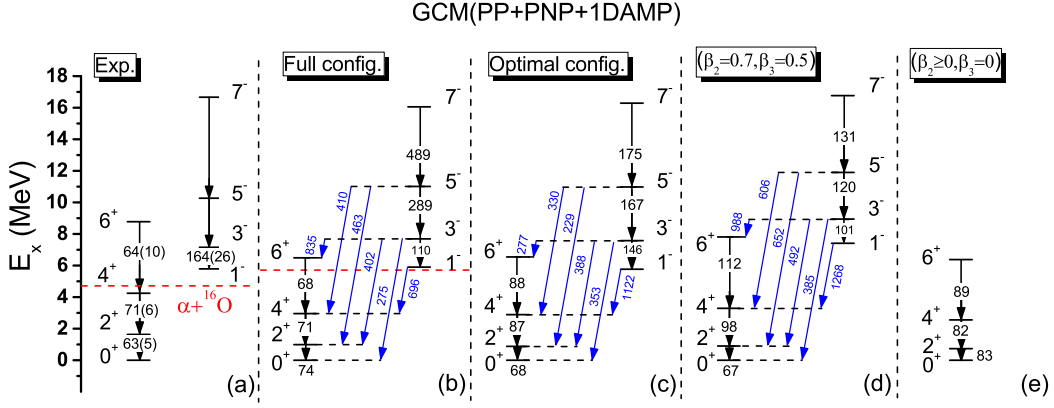}
\caption{\label{pes}(Color online) Low-lying spectra in $^{20}$Ne. The numbers on the arrows are the $E2$ (black color, in $e^2$fm$^4$) and $E3$ (blue color, in $e^3$fm$^6$) transition strengths. The results by mixing different sets of configurations in the GCM calculations are plotted in (b)-(e), respectively. The optimal configurations are those indicated in Fig.~\ref{PES:MF}. The  configuration at ($\beta_2=0.7,\beta_3=0.5$) corresponds to the minimum of $0^+$ energy surface in Fig.~\ref{PES:proj}. The experimental and calculated $\alpha+^{16}$O threshold energies are plotted with horizontal dashed lines in (a) and (b), respectively.}
 \label{Spectra}
\end{figure*}

Figure~\ref{PES:MF} displays the mean-field energy surface in the $(\beta_2,\beta_3)$ plane for $^{20}$Ne. A global energy  minimum is found at the reflection-symmetric prolate shape with deformation parameters $(\beta_2=0.5,\beta_3=0)$. However, in the ground state this equilibrium shape is unstable against fluctuations in octupole-shape. The excitation energy of a possible octupole vibration is expected to be much lower than that of the quadrupole vibration. A schematic picture of nuclear shapes corresponding to some selected configurations along the optimal path on the energy surface is also plotted. It is shown clearly that the reflection-asymmetric diatomic molecular structure ($\alpha+^{16}$O) is progressively developed with increasing values of $\beta_3$ along the optimal path.

Figure~\ref{PES:proj}(a)-(h) show  the projected energy surfaces of the reference states in the $(\beta_2,\beta_3)$ plane. The projections are carried onto the particle numbers $(N, Z)=(10,10)$ and onto different spin-parity values $J^\pi=0^+, 1^-, 2^+, \ldots, 7^-$. For each value $J^\pi$ the projected energy of the reference state $\vert q\rangle$ is given by the diagonal element of the hamiltonian kernel divided by that of normal kernel $E^{J\pi}(q)=\mathscr{H}^{J\pi}_{\kappa, \kappa}/\mathscr{N}^{J\pi}_{\kappa, \kappa}$ with $K=0$. Due to the dynamical correlation energy gained from symmetry restoration, the reflection-asymmetric configurations become favored in energy for both positive- and negative-parity states. On the energy surface with spin-parity $0^+$, there is a minimum with deformation parameters ($\beta_2=0.7,\beta_3=0.5$), which is, however, very soft along both $\beta_2$ and $\beta_3$ directions. With the increasing of spin $J$, the energy minimum becomes more stable but drifts to the configurations with smaller multipole moments. A similar trend is also observed for the energy surfaces with negative parity.

The changes in the projected energy surfaces with increasing spin can be partially related to the deformation-dependent moment of inertia in the sense that a larger moment of inertia gives a lower rotational energy. To shed some light on this issue,  we plot in Fig.~\ref{fig:MoI}(a) the moment of inertia along $x$-axis ${\cal{J}}_x$ in the deformation $(\beta_2,\beta_3)$ plane calculated with Inglis-Belyaev formula~\cite{Belyaev61}
  \begin{eqnarray}
  \label{MOI}
  {\cal{J}}^{IB}_x(\beta_2, \beta_3)=\sum_{m,j}f_{mj}\left|\langle m| \hat J_x|j\rangle\right|^2,
\end{eqnarray}
where $\hat J_x$ is the $x$-component of angular momentum operator. The coefficient between the $m$-th and $j$-th single-(quasi)particle states is defined as $f_{mj}\equiv(u_mv_j-v_mu_j)^2/(E_m+E_j)$ with $E_{k=m, j}=\sqrt{(\epsilon_k-\lambda)^2+f^2_k\Delta^2_k}$ representing quasiparticle energies, $v_k$ for the occupation probability, $\Delta_k$ for the pairing gap, $\epsilon_k$ for the energy of the $k$-th  single-particle state and $f_k$ for the energy-dependent smooth cutoff factor~\cite{Krieger90}. Fig.~\ref{fig:MoI}(a) shows a peak of ${\cal{J}}^{IB}_x$ located around $(0.5, 0.2)$, to which the energy minimum of the projected energy surface drifts with the increase of spin. At the deformation region $\beta_2>0.5$ with pairing collapse the dominant contribution to the moment of inertial comes from the $3/2^+$ and $1/2^+$ orbitals, whose energy difference is increasing steadily with $\beta_2$, resulting in a smooth decreasing of the ${\cal{J}}^{IB}_x$.
It has to be mentioned that the Inglis-Belyaev formula (\ref{MOI}) might be invalid for the states with two-center cluster structures. A further dedicated study of the moments of inertia for such states is required to deepen our understanding on issue.

%%%%%%%%%%%%%%%%%%%%%%%%
\begin{figure}[tb]
\begin{center}
\includegraphics[width=7.5cm]{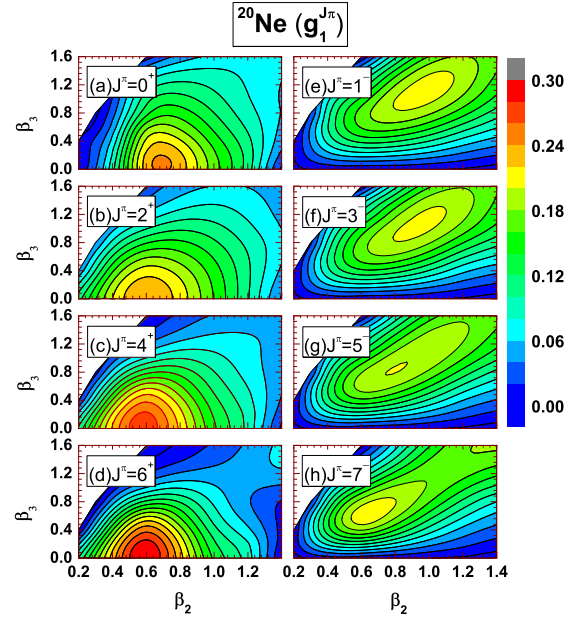}
\caption{(color online) The distribution of collective wave functions $g^{J\pi}_{1} (q_a)$ for the low-lying parity-doublet states ($K^\pi=0^\pm$) of  $^{20}$Ne in $(\beta_2, \beta_3$) plane.} \label{wfs}
\end{center}
\end{figure}
%%%%%%%%%%%%%%%%%%%%%%%%
%%%%%%%%%%%%%%%%%%%%%%%%
\begin{figure}[tb]
\begin{center}
\includegraphics[width=7cm]{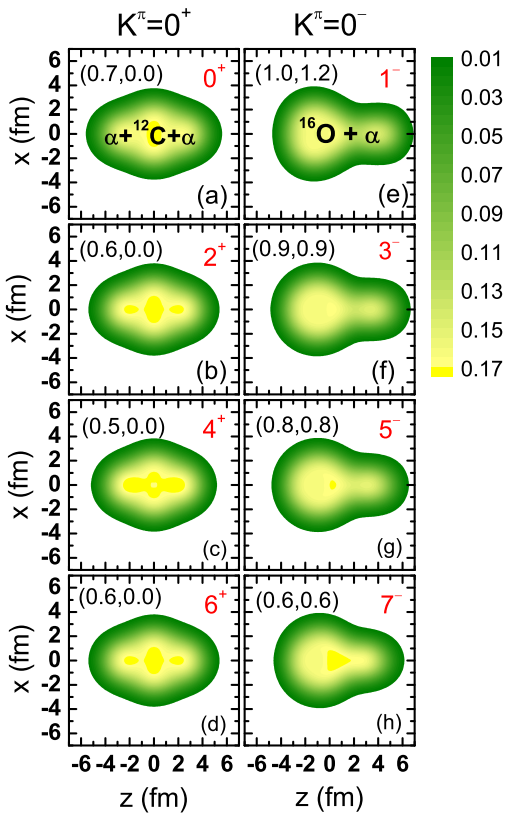}
\end{center}
\caption{(color online) The total density distribution (fm$^{-3}$) of the dominant configuration in the low-lying states of $^{20}$Ne in $x$-$z$ plane.}
\label{dens}
\end{figure}
%%%%%%%%%%%%%%%%%%%%%%%%%%%%%%%%%%%%%

The expectation value of the spin-orbit potential has been used as an indicator of formation or dissociation of cluster structure in the AMD studies~\cite{Kimura04}. Following Ref.~\cite{Yao14-O}, we calculate the spin-orbit interaction energy in two ways,
\bsub \begin{eqnarray}
 \label{SO1}
   SO1
   &=&-\sum_j v^2_j  \langle j\vert V_{\rm s.o.} \vert j\rangle,\\
   \label{SO2}
   SO2
   &=&\sum_j v^2_j  \Big\vert \langle j\vert   V_{\rm  s.o.} \vert j\rangle  \Big\vert
   \end{eqnarray} \esub
 for each referencee state $\vert q\rangle$,  where the ``$-$" sign is introduced in defining $SO1$ to have positive values because the spin-orbit interaction in atomic nuclei is attractive~\cite{Ring96}
   \begin{eqnarray}
    \label{spinorbit}
   \langle j\vert   V_{\rm s.o.} \vert j\rangle  =   \langle j\vert  \dfrac{1}{4m^2}(\nabla V_{\ell s})\cdot (\mathbf{p}\times\sigma)\vert j\rangle,
    \end{eqnarray}
   with $V_{\ell s}=\dfrac{m}{m_{\rm eff.}}(V_0-S)(\bm{r})$, and $m_{\rm eff.} = m-\dfrac{1}{2} (V_0-S)(\bm{r})$, where $V_0(\bm{r})$ and $S(\bm{r})$ are respectively the vector and scalar potentials in Dirac equation for nucleons. $m$ is the bare mass of nucleon. Fig.~\ref{fig:MoI}(c) shows that both $SO1$ and $SO2$ are increasing with the decreasing of quadrupole deformation $\beta_2$ at $\beta_3=0$, consistent with the finding in the AMD studies~\cite{Kimura04} that the cluster structure is being weakened gradually with the increase of spin.

%%%%%%%%%%%%%%%%%%%%%%%%

Figure~\ref{Spectra} displays the low-energy spectra  for $^{20}$Ne from the GCM calculation with different sets of configurations. Both the energy spectrum and the electric multipole transition strengths for low-lying parity-doublet bands are reproduced after taking into account the dynamical octupole vibration effects, as shown in the comparison of Figs.~\ref{Spectra}(b) and (e). In contrast to Fig.~\ref{Spectra}(e) by mixing only the prolate configurations, the full configuration-mixing calculation yields a ground-state rotational band with slightly reduced $E2$ transition strengths and stretched spectrum, which are closer to data. In particular, the negative-parity band is also  reproduced reasonably. Figure~\ref{Spectra}(d) shows that the single quadrupole-octupole deformed configuration  ($\beta_2=0.7,\beta_3=0.5$)  of the energy-minimum on the projected $0^+$ energy surface is able to reproduce the main feature of the spectrum. The results (except for the E2 transition strengths in the negative-parity band) by mixing the configurations along the optimal path, cf. Fig.~\ref{Spectra}(c), are in between those by mixing the full configurations and by the single configuration at ($\beta_2=0.7,\beta_3=0.5$).

%The energy displacement between the positive and negative parity bands is determined by the norm overlap $\langle q\vert \hat P\vert q\rangle$ where $\hat{P}$ the parity operator. In light nuclei, because of the small number of particles, this quantity decreases very slowly with increasing $\beta_3$ and is still nonzero for ($\beta_2=0.7,\beta_3=0.5$) in $^{20}$Ne. This situation is different from the case in heavy nuclei such as $^{224}$Ra, where the norm overlap decreases to zero quickly with the increase of $\beta_3$~\cite{Yao15}. As a result, based on this single-reference state, we obtain a large energy displacement between the positive and negative parity bands, as shown in Fig.~\ref{Spectra}(d). In other words, the parity splitting is more or less reproduced by the projection calculation based on one single configuration of the energy-minimum on the $0^+$ energy surface.

Figure~\ref{wfs} displays the distributions of  collective wave functions $g^{J\pi}_\alpha (q) \equiv \sum_{q'} \left[\mathscr{N}^{J\pi}_{q, q'}\right]^{1/2}f^{J\pi\alpha}_{q'}$ in the deformation $(\beta_2,\beta_3)$ plane, \jmy{where the norm kernel $\mathscr{N}^{J\pi}_{q, q'}$ has been defined in Eq.(\ref{kernel}) with $K=0$. The way to calculate the $g^{J\pi}_\alpha (q)$ can be found in Refs.~\cite{Ring80,Yao10}. It has to be stressed that the $g^{J\pi}_\alpha (q)$ is different from the quantity $\tilde g^{J\pi}_\alpha (q) \equiv \langle q \vert  J^\pi NZ; \alpha\rangle$, the module square of which represents the probability to find the mean-field configuration $\vert q\rangle$ in the GCM state $\vert J^\pi NZ; \alpha\rangle$. However, the quantity $g^{J\pi}_\alpha (q)$ is orthonormal to each other and is usually adopted to analyze the predominant configuration in the GCM wave functions~\cite{Yao10,Rodriguez10,Yao13}. It is shown in Fig.~\ref{wfs} that the collective wave functions $g^{J\pi}_\alpha (q)$ of the states in the $K^{\pi}=0^+$ band are peaked in a reflection-symmetric prolate configuration around $(0.5, 0.0)$. With increasing angular momentum, the collective wave functions becomes increasingly concentrated, indicating the stabilization of nuclear shape under rotation. In contrast, we find a relatively broad distribution of the collective wave functions for the states in the $K^{\pi}=0^-$ band. The peak position drifts towards a smaller deformed region with increasing angular momentum.}

% despite that the projected potential energy surfaces in Fig.~\ref{PES:proj} exhibit minima with nonzero $\beta_3$ value. In other words, the dominant configuration does not correspond to the configuration with the lowest diagonal element of the energy kernel. This phenomenon is very common in light nuclei as observed in $^{16,18}$C~\cite{Yao01-C} and $^{24}$Mg~\cite{Yao10}, where the off-diagonal elements are comparable in size to the diagonal elements.  Additionally,

 The total nucleon densities in $y$-$z$ plane corresponding to the dominant configurations in each spin-parity state are displayed in Fig.~\ref{dens}. The positive-parity states have probably a compact $\alpha+^{12}$C+$\alpha$ structure, consistent with the recent mean-field studies based on the relativistic DD-ME2 force~\cite{Ebran12}, while the negative-parity states have  a reflection-asymmetric diatomic $^{16}$O+$\alpha$ molecular structure.  The broad distribution of the collective wave functions of the negative-parity states may indicate the nonlocalization of the cluster structures~\cite{Zhou13}. In the dominant configuration, the distance between the $\alpha$ cluster and the $^{16}$O cluster becomes smaller as the angular momentum increases and the clustering feature becomes weaker. This has been found also in AMD calculations, where this phenomenon was interpreted as the consequence of an increasing spin-orbit interaction~\cite{Kimura04}. A quite similar phenomenon was also discussed with the $5\alpha$ GCM calculation~\cite{Nemoto75}.

% On the other hand, as we have discussed in Fig.~\ref{fig:MoI},  the shift of the dominant configuration from  larger to smaller deformations with the increasing spin is partially driven by the deformation-dependent moment of inertia, which is peaked around $\beta_2\approx0.5,\beta_3\approx0.2$.

 In summary, we have presented a beyond-mean-field study of reflection-asymmetric molecular structures in $^{20}$Ne for the first time in the framework of multireference relativistic energy density functional theory. The dynamical correlations related to the restoration of symmetries broken at the mean-field level and to shape fluctuations are taken into account within the quantum-number projected generator coordinate method. Both the energy spectrum and the electric multipole transition strengths for low-lying parity-doublet bands have been reproduced after taking into account the effect of dynamical octupole vibrations.  We conjecture that  the special deformation-dependent moment of inertia governed by the underlying shell structure is responsible for the rotation-induced dissolution of the $\alpha+^{16}$O molecular structure in the negative-parity states. However, a further dedicated study of the deformation-dependent moment of inertia is required to address this issue. The present study also provides a starting point to investigate the dynamics of nuclear $\alpha$-decay in a full microscopical way in the future.

%\section*{acknowledgements}
%{\em Acknowledgements.}$-$.
One of the authors (E.F.Z.) thanks T. Nik\u{s}i\'{c} and D. Vretenar for their hospitality during his visit to University of Zagreb. We are grateful to K. Hagino, H. Mei, T. Nik\u{s}i\'{c}, G. Scamps and D. Vretenar for fruitful discussions and thank the anonymous referee for the constructive suggestions. This work was partially supported by the Major State 973 Program 2013CB834400,  the NSFC under Grant Nos. 11575148, 11475140, 11335002, and 11305134, by the DFG cluster of excellence \textquotedblleft Origin and Structure of the Universe\textquotedblright (www.universe-cluster.de), and by the Chinese-Croatian project
``Universal models of exotic nuclear structure".


\begin{thebibliography}{9}
\bibitem{Ikeda80}   K. Ikeda, H. Horiuchi, and S. Saito, Prog. Theor. Phys. Suppl. 68, 1 (1980).
\bibitem{Fynbo05}   Hans O. U. Fynbo et al., Nature 433, 13 (2005).
\bibitem{Oertzen06}  W. von Oertzen, M. Freer, and Y. Kanada-En$'$yo, Phys. Rep. 432, 43 (2006).
\bibitem{Freer07}   M. Freer, Rep. Prog. Phys. 70, 2149 (2007).
\bibitem{Kanada12}  Y. Kanada-En'yo, M. Kimura, and A. Ono, Prog. Theor. Exp. Phys. 01A202 (2012).
\bibitem{Typel10}   S. Typel, G. R\"{o}pke, T. Kl\"{a}hn, D. Blaschke, and H. H. Wolter, Phys. Rev. C 81, 015803 (2010).

 \bibitem{Ikeda68}  K. Ikeda, N. Takigawa and H. Horiuchi, Prog. Theor. Phys. Suppl. E68, 464 (1968);
                    H. Horiuchi, K. Ikeda and Y. Suzuki, Prog. Theor. Phys. Suppl. 52, 89 (1972).

 \bibitem{Horiuchi12} H. Horiuchi, K. Ikeda, and K. Kato, Prog. Theor. Phys. Suppl. 192, 1 (2012).
\bibitem{Bayman58} B. F. Bayman and A. Bohr, Nucl. Phys. 9, 596 (1958).

\bibitem{Wiringa00} R. B. Wiringa  and Steven C. Pieper, J. Carlson, V. R. Pandharipande, Phys. Rev. C 62, 014001 (2000).
\bibitem{Chernykh07} M. Chernykh, H. Feldmeier, T. Neff, P. von Neumann-Cosel and A. Richter, Phys. Rev. Lett. 98, 032501 (2007).

 \bibitem{Lee72} H. C. Lee and R. Y. Cusson, Phys. Rev. Lett. 29, 1525 (1972); Phys. Rev. Lett. 30, 153 (1973).
  \bibitem{Matsuse75} T. Matsuse, M. Kamimura and Y. Fukushima, Prog. Theor. Phys. 53, 706 (1975).
  \bibitem{Nemoto75} F. Nemoto, Y. Yamamoto, H. Horiuchi, Y. Suzuki and K. Ikeda, Prog. Theor. Phys. 54, 104 (1975).
   \bibitem{Marcos83} S. Marcos, H. Flocard, P.H. Heenen, Nucl. Phys. A410, 125 (1983).
\bibitem{Kanada95} Y. Kanada-En'yo and H. Horiuchi, Prog. Theor. Phys. 93, 115 (1995).
\bibitem{Kimura04} M. Kimura, Phys. Rev. C 69, 044319 (2004).
\bibitem{Taniguchi04} Y. Taniguchi, M. Kimura and H. Horiuchi, Prog. Theor. Phys. 112, 475 (2004).
   \bibitem{Ohta04} H. Ohta, K. Yabana, T. Nakatsukasa, Phys. Rev. C 70, 014301 (2004).

\bibitem{Ebran12} J. P. Ebran, E. Khan, T. Niksic, and D. Vretenar, Nature (London) 487, 341 (2012);
Phys. Rev. C 89, 031303(R) (2014).
\bibitem{Arumugam05} P. Arumugam,  B. K. Sharma,  S. K. Patra, Phys. Rev. C 71, 064308 (2005).
\bibitem{Zhang10} W. Zhang, H.-Z. Liang, S.-Q. Zhang, and J. Meng, Chin. Phys. Lett. 27, 102103 (2010).
\bibitem{Ichikawa11} T. Ichikawa, J. A. Maruhn, N. Itagaki, and S. Ohkubo, Phys. Rev. Lett. 107, 112501 (2011).
\bibitem{Zhao15} P. W. Zhao, N. Itagaki, and J. Meng, Phys. Rev. Lett. 115, 022501  (2015).
\bibitem{Girod13} M. Girod and P. Schuck, Phys. Rev. Lett. 111, 132503 (2013).
\bibitem{Ring80} P. Ring and P. Schuck, The Nuclear Many-Body Problem (Springer-Verlag, Berlin, 1980).
\bibitem{Duguet14} T. Duguet, The Euroschool on Exotic Beams, Vol. IV Lecture Notes in Physics, 879,  293 (2014).
\bibitem{Yao14-O} J. M. Yao, N. Itagaki, and J. Meng, Phys. Rev. C 90, 054307 (2014).
\bibitem{Hill57} D. L. Hill and J. A. Wheeler, Phys. Rev. 89, 1102 (1953);
                 J. J. Griffin and J. A. Wheeler,  Phys. Rev.  108, 311 (1957).
\bibitem{Zhao10} P. W. Zhao, Z. P. Li, J. M. Yao, and J. Meng, Phys. Rev. C 82, 054319 (2010).
\bibitem{Gambhir90} Y. K. Gambhir, P. Ring, and A. Thimet, Ann. Phys. (N.Y.) 198, 132 (1990).
\bibitem{Krieger90} S. J. Krieger, P. Bonche, H. Flocard, P. Quentin, and M. S. Weiss, Nucl. Phys. A 517, 275 (1990).
\bibitem{Bonche90} P. Bonche, J. Dobaczewski, H. Flocard, P.-H. Heenen, and J. Meyer, Nucl. Phys. A 510, 466 (1990).
\bibitem{Robledo09} L. M. Robledo, Phys. Rev. C 79, 021302 (2009).
\bibitem{Bertsch12} G. F. Bertsch and L. M. Robledo, Phys. Rev. Lett. 108, 042505 (2012).
\bibitem{Yao10} J. M. Yao, J. Meng, P. Ring, and D. Vretenar, Phys. Rev. C 81, 044311 (2010).
\bibitem{Yao14} J. M. Yao, K. Hagino, Z. P. Li, J. Meng, and P. Ring, Phys. Rev. C 89, 054306 (2014).
\bibitem{Yao15} J. M. Yao, E. F. Zhou, and Z. P. Li, Phys. Rev. C 92, 041304 (R) (2015).
\bibitem{Belyaev61} S. T. Belyaev, Nucl. Phys. 24, 322 (1961).
 \bibitem{Ring96} P. Ring, Prog. Part. Nucl. Phys. 37, 193 (1996).
 \bibitem{Rodriguez10} T. R. Rodriguez and J. L. Egido, Phys. Rev. C 81, 064323 (2010).
\bibitem{Yao13} J. M. Yao,  M. Bender, and P.-H. Heenen, Phys. Rev. C 87, 034322 (2013).
\bibitem{Zhou13} B. Zhou, Y. Funaki, H. Horiuchi, Z. Z. Ren, G. R\"{o}pke, P. Schuck, A. Tohsaki, C. Xu, and T. Yamada, Phys. Rev. Lett. 110, 262501 (2013).

\end{thebibliography}
\end{document}